# Earthquakes and the wealth of nations: The cases of Chile and New Zealand

Diego Díaz*  Pablo Paniagua†  Cristián Larroulet*

**Draft paper: version May 20th, 2024**

**Abstract**

The consequences of natural disasters, such as earthquakes, are evident: death, coordination problems, destruction of infrastructure, and displacement of population. However, according to empirical research, the impact of a natural disaster on economic activity is mixed. Natural disasters could have significant economic effects, especially in developing economies. This is particularly important for highly seismic countries such as Chile and New Zealand. This paper contributes to the literature on natural disasters and economic development by analyzing the cases of two affected regions within these countries in the wake of major earthquakes experienced during 2010-2011: Maule (Chile) and Canterbury (New Zealand). We examine the impact of natural disasters on GDP per capita by applying the synthetic control method. Using the synthetic approach, we assess the effects of these two earthquakes by building counterfactuals to compare their recovery trajectories. We find that Chile and New Zealand experienced opposite economic effects. The Canterbury region grew 10% more in three years than its synthetic counterfactual without the earthquake, while the Maule region declined by 5%. We build synthetic controls at a regional and economic-sector level, looking at aggregated and sectoral effects. The difference in institutions, such as property rights and the large amount of government spending given for reconstruction after the New Zealand earthquake relative to Chile's, help to explain the difference in outcomes.

* School of Business and Economics (FEN), Universidad del Desarrollo (UDD), Santiago.
† King's College London, London, UK. Corresponding author: ppaniagua@udd.cl.

*"Earthquakes are the Earth's way of freeing itself from its ghosts."*
*Alberto Fuguet*

1. **Introduction**

At 3.34 A.M. on Saturday, February 27th, 2010, an earthquake of magnitude 8.8 Mw hit the region of Maule in Chile. It has been registered as the sixth largest earthquake ever recorded in human history according to the USGS Earthquake Hazards Program (USGS, 2019), and the second strongest to hit Chile (Bárcena et al., 2010). The sudden vertical movement of the ocean floor also caused a tsunami along the fault-rupture area, devastating broad coastal zones, cities, and towns of central Chile. The quake left 562 people dead and produced economic costs estimated at US$ 30 billion (EMDAT, 2019)—approximately 18% of the Chilean GDP in 2010.

Analogously, in the same year, and approximately 9.158 kilometers west, at 4:35 A.M. on September 4th, 2010, a 7.1 Mw earthquake struck New Zealand in Canterbury. Although the economic damage was extensive (estimated at around US$ 6.5 billion), there were no casualties, in part because of the location (the disaster took place near the small town of Darfield) and the time of the day when it happened (at night). However, several aftershocks occurred during 2011, which had severe consequences. The most intense occurred on February 22nd, 2011, at 12:51 P.M. in the highly populated city of Christchurch in the same region. Although of magnitude 6.3 Mw, the aftershock had more intense and devastating oscillations. The severe effects can be explained, at least in part, because many of the buildings' structures were already compromised by the stronger, earlier event. One hundred eighty-five people died, and economic damage was estimated at US$ 18 billion (EMDAT, 2019), which accounted for 10% of New Zealand's GDP in 2011. From these two experiences, closely related to each other in terms of distance and magnitude, we can grasp and assess the economic consequences of natural disasters such as earthquakes on the wealth and development of nations. This will be the task of this paper.

The immediate consequences of natural disasters, such as earthquakes, are evident: death, coordination problems, destruction of infrastructure, and displacement of population. However, the impact of a natural disaster on economic activity is mixed according to empirical research, and much dependent upon factors such as level of development and state capacity (Cavallo, et al., 2013; Cavallo and Noy, 2011; Raddatz, 2009; Toya and Skidmore, 2007; Kellenberg and Mobarak, 2011). The economic literature is inconclusive concerning the role that natural disasters could play in the long run on the wealth of nations (duPont IV and Noy, 2015). Natural disasters could have significant and long-lasting economic effects, especially in developing economies (Best and Burke, 2019; Cavallo et al., 2010; Toya and Skidmore, 2007; Lian et al., 2022). This is important for highly seismic countries such as Chile and New Zealand.

This paper contributes to the literature on natural disasters and economic development by analyzing the cases of these two countries in the weak of major earthquakes experienced during 2010-2011. We examine the impact of natural disasters on GDP per capita by applying the synthetic control method (Abadie, 2021; Abadie et al., 2015; Rayamajhee et al., 2024). Using the synthetic approach, we assess the effects of these two earthquakes by building counterfactuals to compare the recovery trajectory of these cases. In this paper, we evaluate the economic impact of the two previously mentioned earthquakes: Chile in 2010 and New Zealand in 2011. We decided to study the Christchurch earthquake instead of the Darfield one in 2010 in New Zealand, given that it was the most destructive one in terms of economic damage and it most likely had a more significant impact in the long term.

Our synthetic controls suggest that Chile and New Zealand experienced opposite economic effects: We find that the Canterbury region (New Zealand) grew 10% more in three years than its synthetic counterfactual without the earthquake, while the Maule region (Chile) declined in economic activity by 5% compared to its counterfactual. We build synthetic controls at a regional level, looking also at aggregated and sectoral effects. The difference in institutions, such as property rights and the large amount of government spending given for reconstruction after the New Zealand earthquake relative to Chile's, help to explain the difference in outcomes. This analysis is relevant for the literature on development economics and disasters since the Chilean and New Zealand cases are important regarding quake magnitude, yet severely disregarded by the literature. However, despite their severity, they have been largely unexplored from an empirical perspective (Miller, 2014). Thus, the literature has ignored these cases without delving deeper into causality and counterfactuals to make an in-depth analysis. Closing this gap in the literature is the main contribution of this paper using the synthetic control method (SCM).

A short-term quantitative and economic assessment of the Canterbury region was made by researchers from the Central Bank of New Zealand (Parker & Steenkamp, 2012; see also Miller, 2014). They concluded that, although the impact of the 2011 Canterbury earthquake was severe (10% of national GDP), the Canterbury economy was resilient, and the broader New Zealand economy was unaffected. They also noticed some indirect socioeconomic effects, such as a net outflow of residents from Christchurch to other regions and overseas destinations. Some sectors, such as retail and accommodation, were also more negatively affected than others. Other economic areas, like the construction sector, saw their demand increase sharply following the earthquake. Paradoxically, the Chilean natural disaster, despite being one of the most significant quakes in recorded human history (Cárdenas-Jirón, 2013), has been severely understudied by social scientists using empirical techniques (Bárcena et al., 2010; Dussaillant and Guzmán, 2014).

Today, over ten years after both events took place, we provide an economic assessment of the consequences of these two earthquakes, looking at the short and long-term impact using synthetic counterfactuals. To the best of our knowledge, this is the first article that attempts to do this simultaneously for the earthquakes in Chile and New Zealand. Our approach consists in estimating a counterfactual of the affected regions in each country by the synthetic control method (SCM), a technique used in comparative case studies (Abadie and Gardeazabal, 2003; Abadie et al., 2020; Absher et al., 2020; Gilchrist, et al., 2023). The SCM has been used extensively in economics and political science in the last decades, including in natural disasters' damage assessment studies (Barone and Mocetti, 2014; duPont IV et al., 2015; Lynham et al., 2017; Rayamajhee et al., 2024; Lian et al., 2022).

Recently, several articles have been published that look at specific earthquakes and other natural disasters, assessing their economic impact through the SCM in places such as Italy, Japan, Haiti, and Hawaii (Best and Burke, 2019; Cerqua and Di Pietro, 2017; duPont IV, et al., 2015; duPont IV and Noy, 2015; Barone and Mocetti, 2014; Lynham, et al., 2017). Both duPont IV's articles study the impact of the Kobe earthquake in Japan in 1995, with different datasets. Barone and Mocetti (2014) studied two earthquakes in different Italian regions: One in Friuli in 1976 and one in Irpinia in 1980. Best and Burke (2019) studied the impact of the devastating earthquake in Haiti in 2010, which took place a month and a half earlier than the Chilean earthquake in the same year. Cerqua and Di Pietro (2017) determine the impact of the Abruzzi (Italy) earthquake 2009. Unlike the other articles, which study the economic impact, they look at university enrollment, assessing the effect of the disaster on enrollment at the University of L'Aquila. Of the articles mentioned, except for Barone and Mocetti (2014) and Cerqua and Di Pietro (2017), all find that earthquakes have a negative and long-lasting impact on the economy and development. These findings seem to be validated by Lian et al. (2022), who quantify the dynamic effects of natural disasters on real GDP per capita for many episodes using a synthetic control approach. The authors find a "persistently large deviation of real GDP per capita from the counterfactual trend exists five years after a severe shock in many countries" (ibid., 1). This paper contributes to such literature by focusing on the two neglected cases of Chile and New Zealand.

The paper proceeds as follows. Section 2 provides an in-depth literature review of the intersection between disasters and development economics, discussing the impact of natural disasters on output and economic growth. Section 3 describes the economies of Chile and New Zealand, comparing their socioeconomic characteristics, leading economic indicators, and the two earthquakes. Section 4 develops the synthetic control approach and reviews the data and method employed: using the SCM, we construct a synthetic or counterfactual for the economic performance of the affected regions of Chile and New Zealand *in the absence* of their devastating earthquakes. Section 5 presents the results of both case studies and provides some explanations considering the economic literature concerning the role of institutions and state capacity. Section 6 concludes.

2. **Literature review**

Earthquakes and other natural disasters, despite being catastrophic events, at least provide an opportunity for social scientist to study their impact on economic growth, institutions, and other socioeconomic and macroeconomic indicators (Hallegatte and Przyluski, 2010; Okuyama, 2007; Kahn, 2005; Cohen, et al., 2008). Since they cannot be predicted with current technology, quakes' full short-term consequences are felt immediately in the affected areas (Cavallo et al., 2013). The synthetic control method (SCM) (Abadie, 2021; Abadie and Gardeazabal, 2003; Abadie et al., 2020; Absher et al., 2020; Gilchrist et al., 2023; Spruk, 2019) is an ideal empirical tool for estimating the economic impact of unpredictable natural disasters since it allows to create a reasonable counterfactual of the affected unit and determine what would have happened *had the event not taken place*, even in the long term (Lian et al., 2022; duPont IV and Noy, 2015).

The literature related to natural disasters is vast. For instance, a quick search on the Web of Science database with the topic "natural disasters" will produce more than 6.160 published papers. Adding "economic growth" will make the results more manageable, reducing the sample to 123 articles. It is not our intention to produce a comprehensive review on this subject, since there are already published reviews (Shabnam, 2014; Felbermayr and Gröschl, 2013, 2014; Cavallo and Noy, 2011; Chhibber and Laajaj, 2008; Okuyama, 2003; Díaz and Larroulet, 2021). Such empirical reviews converge on one puzzling idea, which is that the current state of the literature regarding natural disasters and economic growth in the long run is inconclusive since other relevant factors such as institutional quality, trade openness, and state capacity could mitigate the adverse effects of natural shocks (Felbermayr and Gröschl, 2014; Noy, 2009).

For instance, Cavallo et al (2013, 1549) argue that after controlling for ex-post-political factors, "even extremely large disasters do not display any significant effect on economic growth". Toya and Skidmore (2007, 20) suggest that "countries with higher income, higher educational attainment, greater openness, more complete financial systems and smaller government experience fewer [human and economic] losses". Similarly, Noy (2009) argues that the severity of the macroeconomic output costs depends on the level of development: "developing countries, and smaller economies, face much larger output declines following a disaster of similar relative magnitude than do developed countries or bigger economies". The author finds that countries with better institutions, higher per capita income, higher degree of openness to trade, and higher levels of government spending withstand disasters better, preventing spillovers into the macro-economy (ibid., 221).

Recently, to deal with this ambiguity in the results of the literature, several case studies about natural disasters have used the SCM to determine the effects of disasters on different variables, such as income, population, production levels, number of employers, and unemployment rate (Best and Burke, 2019; Coffman and Noy, 2012; Lynham et al., 2017; Barone and Mocetti, 2014; Rayamajhee et al., 2024; duPont IV and Noy, 2015; duPont IV et al., 2015; Cerqua and Di Pietro, 2017). As unlikely as it may sound, a natural disaster might have some positive effects on the economy, as some empirical studies have suggested. For instance, in some cases disasters could increase economic output (Albala-Bertrand, 1993; Skidmore and Toya, 2002; Leiter et al, 2009; Loayza et al, 2012; Fomby et al., 2013; Felbermayr and Gröschl, 2013). On the other hand, there has not been a lack of studies also showing that disasters have only neutral or negative effects on output (Rasmussen, 2004; Noy, 2009; Raddatz, 2007, 2009; Toya and Skidmore, 2007; Hochrainer, 2009; Schumacher and Strobl, 2011; Cavallo et al., 2013; Felbermayr and Gröschl, 2014).

Many of these recent articles rely on the Emergency Events Database EM-DAT, which provides worldwide coverage from the Centre for Research on the Epidemiology of Disasters (CRED) at Louvain's Catholic University. The database registers Hydro-meteorological, geophysical, and biological disasters if at least one of four conditions is met: i) At least ten people must perish by the disaster, ii) 100 people must be affected by the disaster, iii) a state of emergency is declared, and iv) international assistance is requested (Cavallo et al, 2013). Yet, Felbermayr and Gröschl (2014) note that the EM-DAT database could present problems for growth regressions for a couple of reasons: First, the monetary damage of a disaster is higher in a richer economy, as can be noted given that the database measure of disaster intensity correlates with GDP per capita. Second, insurance coverage may also correlate with GDP per capita. When more insurance claims are filed in an affected area, it is more likely that a natural disaster will be reported and, hence, more likely that the criteria for inclusion will be met. In other words, the probability of inclusion might be correlated with GDP per capita, and both problems individually lead to an upward bias in growth regressions of disasters on income.

Whether natural disasters are favorable or unfavorable for economic growth is still an open debate in the literature (Cavallo et al., 2013; Cavallo and Noy, 2011; Raddatz, 2009; Toya and Skidmore, 2007; Kellenberg and Mobarak, 2011). Although standard economic growth theory predicts that the impact should be negative, there is also a theoretical precedent for positive effects in endogenous growth models, as was noted by Aghion and Howitt (1998) and Howitt and Aghion (1998), in a Schumpeterian growth model in which positive technological change after a disaster could lead to higher growth because of capital replacement. The mechanism commonly mentioned that suggests disasters could have positive effects (Albala-Bertrand, 1993; Benson & Clay, 2004; Okuyama, 2003; Stewart and Fitzgerald, 2001) has also been studied in theoretical terms more recently by Hallegate and Dumas (2009), who call the mechanism the "productivity effect". According to these authors: "disasters might have positive economic consequences, through the accelerated replacement of capital. This possibility is referred to as the productivity effect" (ibid., 777). They find that "with or without productivity effect, short-term constraints on reconstruction have a large influence on the deepness and duration of the negative consequences of a disaster. Over the long-term, these constraints do not play any role and disasters do not influence the long-term growth rate, unless the capacity to fund and carry out the reconstruction is lower than a threshold value, related to the intensity and frequency of disasters" (ibid., 784).

Shifting our attention to specific articles that have studied the impact of natural disasters by the SCM, we searched published papers in the Web of Science core collection, looking for articles that match the topic search of natural disasters and synthetic control. We found 12 relevant articles with both terms present in the article's title, abstract, or keywords (see Table A1 in the appendix). Of these articles, we have selected only those focusing on earthquakes, reducing the number of studies to 5, as depicted in Table 1.

All five articles that have made a detailed case study of an earthquake using the SCM, except for Cerqua and Di Pietro (2017), are similar to ours in the sense that they all estimate the economic impact of an earthquake using macroeconomic data. The variables they study, among others, are GDP per capita (or income) and population, except for Barone and Mocetti (2014), which focuses only on GDP per capita. The Barone and Mocetti (2014) (from now on BM) article has the most in common with ours. Besides the methodology, our work shares a similar objective: to study whether the quality of institutions in a given region affects subsequent growth in GDP per capita following an earthquake using counterfactuals. BM performed a case study of two Italian regions–one north (Friuli) and one south (Irpinia). The authors find that the northern region *grew faster than without the earthquake*, while the opposite is true for the southern region. Interestingly, we obtain similar results for New Zealand and Chile, respectively (see section 5).

When comparing the north region (Friuli) with the south region (Irpinia), the BM study also suggests that the difference occurs due to institutional differences between the two regions, with negative outcomes "more likely to occur in regions with lower pre-quake institutional quality" (ibid). The negative outcome to which the authors refer is a lower level of GDP per capita in the long term.

**Table 1: Literature on earthquakes with the synthetic control method**

| Article | Location and year of earthquake/s | Unit of study | Variable/s of interest | Data | Major conclusions |
|---|---|---|---|---|---|
| Best & Burke (2019) | Léogâne, Ouest department, Haiti (2010) | Country | Several variables* | 13 economic and demographic variables in total. Yearly data for 22 low-income countries from 2004 to 2014. Data source: World Bank WDI, UN, IEA, IMF. | GDP declined 12% on average until 2015, total loss of US$6 billion. Aggregate consumption increased in the short term; investment declined. GDP effect seems persistent. |
| Cerqua & Di Pietro (2017) | L'Aquila, Abruzzi region, Italy (2009) | University | University enrollment | 8 educational, economic, and labor market variables covering different intervals of the period 2000-2012 of 25 Italian universities. Economic and labor market variables are taken from the province where the university is located. Data source: MIUR and ISTAT. | No significant in overall enrollment, but the earthquake changed the composition of the student body. Enrollment of young students decreased while it increased for older students |
| duPont et al. (2015) | Kobe, Hyōgo Prefecture, Japan (1995) | City | Income and population | 67 variables (demographic, environmental, economic, governmental, labor, and spatial-economic) of 1719 cities during the period 1980-2010 at different frequencies (1, 3, and 5 years for different variables). Source: Japanese government's Statistics Bureau. | Incomes and, to a lesser extent, population decreased after the disaster. Persistent effect. |
| duPont & Noy (2015) | Kobe, Hyōgo Prefecture, Japan (1995) | Prefecture | Population and GDP per capita | 7 economic and demographic variables of 20 prefectures covering the period 1975-2009. Source: Japanese government's Statistics Bureau | Population decreases temporarily, returning to trend after 5 years. GDP per capita increased immediately after (perhaps because of the population change) and decreased afterward. The decrease was persistent |
| Barone & Mocetti (2014) | Gemona del Friuli, Friuli region (1976) & Conza, Irpinia region (1980), Italy | Region | GDP per capita | 13 economic, demographic, and institutional variables of 20 Italian regions covering different intervals during the period 1951-2009. Source: ISTAT, AJS, RdB. | Mixed and persistent long-term effects in GDP per capita. The region with better institutional variables (less corruption, higher electoral turnout and newspaper readership) experiences the positive outcome |

* Best and Burke (2019) estimate the impact on 16 variables: GDP, GDP per capita, GDP (sectoral), consumption, gross capital formation, government revenue, government expenditure, imports, exports, external balance, development aid, inflation, personal remittances, electricity production, transport energy use, and population in the largest city.

The BM's results for the northern region (Friuli) are especially interesting given that it is the only case study of an earthquake analyzed with the SCM finding that a good outcome could occur following the disaster (in terms of GDP per capita). Our findings are consistent with Barone and Mocetti (2014), contributing to the literature on the economic consequences of disasters using the SCM. BM constructs an "overall institutional quality index." Therefore, they suggest that the Friuli region obtains higher income levels after the quake because of more robust institutions and financial aid. Their claim is based on the fact that Friuli has on average, lower corruption, fewer politicians involved in scandals, higher turnout, and higher newspaper readership. Following the BM's study, we shed light on why there are differences in outcomes when earthquakes occur in different regions with different levels of institutional development. In our research, we look at earthquakes in different countries. One occurs in Chile, a middle-income country from South America, and the other in New Zealand, a high-income country from Oceania.

Analogous to BM and duPont and Noy (2015), our analysis is made at the regional level, or in the words of BM, with a "within country perspective". Since earthquakes, unlike hydro-meteorological disasters, usually affect a *limited location* within the country, a regional approach is much more likely to produce a reliable and accurate counterfactual when using the SCM compared to nation-level analysis. An alternative may be analyzing at the city level, as in duPont et al. (2015). This could be potentially more precise in measuring impact, but only if the areas selected are correctly identified to avoid influencing the control group (Abadie, 2021). Alas, employing the synthetic control approach is dependent on data availability. Therefore, for the cases of Chile and New Zealand, there are no sources available with local data of economic aggregates at the city level that would permit us to apply the SCM in a different level of aggregation. Another alternative, made recently by Best and Burke (2019) for the Haiti earthquake of 2010, applies the method at the country level. Although this might be the only option for Haiti, given data availability or because of the size and geography of the country since it is possible that the whole country was affected by the disaster; working with smaller and regional units will undoubtedly produce more precise estimates regarding the impact of the treatment (Abadie and Gardeazabal, 2003).

Finally, our reasons for looking at the earthquakes of Chile in 2010 and New Zealand in 2010-2011 are twofold: First, the disasters under study took place only within one year of each other, from February 27th, 2010, to February 22nd, 2011. This makes it more unlikely that variables such as technological change or international economic shocks are causing the differences in outcomes. Second, there are significant institutional differences between both economies, and more so than we can find between different regions within the same country. Third, and finally, these two countries have reliable regional data, and given that natural disasters in those countries usually affect regions very differently, the SCM at a regional level could provide a more accurate picture of the role of earthquakes on the wealth of nations. In section 3, we describe in detail some of the fundamental characteristics of Chile's and New Zealand's economies, suggesting a few mechanisms that might cause the difference in outcomes after each earthquake.

### 3. The cases of Chile and New Zealand: an overview

Chile is geographically on the edge of the oceanic Nazca Plate and the continental South American Plate through most of its territory (Bárcena et al., 2010). New Zealand extends across the Pacific Plate and the Australian Plate, with the North Island located entirely on the Australian Plate and the South Island extending across both plates (Miller, 2014). Since both countries have a significant portion of land near tectonic fault lines, they have above-average volcanic and seismic activity. Regarding income, New Zealand is a highly developed nation that ranks 16th in the Human Development Index and had a GDP per capita of $41,966 US in 2018. On the other hand, Chile's GDP per capita is $15,923 US, and the country ranks 44th in the Human Development Index, and thus among the middle-income range of countries (World Bank, 2019).

The two earthquakes under study were the most destructive ones in Chile and New Zealand's recent history regarding economic damage (Dussaillant and Guzmán, 2014; Parker and Steenkamp, 2012). The one in the Maule region was the second strongest in Chile's history and produced strong tsunamis responsible for most of the 562 casualties. The tsunami alert after the Chilean earthquake was massive, extending to 53 countries, including New Zealand. It was estimated that around 400,000 houses were affected throughout the country, and a state of emergency was declared in Chile (Bárcena et al., 2010). The most affected regions of Maule and Bio-Bio also declared a state of catastrophe. Notably, such a big earthquake that caused many repercussions in Chile and around the world has not been studied at depth by social scientists from an empirical perspective. This paper seeks to remedy this gap in the literature. Paradoxically, the Canterbury earthquake, although significantly lower in magnitude than the one in Maule, has been studied in much greater detail (Doyle and Noy, 2013; Parker and Steenkamp, 2012; Potter et al., 2015; Miller, 2014). In terms of damage, it was also significantly lower: 185 people died, and most of them (115) died due to the collapse of the Canterbury Television (CTV) building (Potter et al., 2015). Although we focus on the economic impact of the earthquakes, the effect is not limited to economics alone, or at least it does not only affect economic sectors directly. Potter et al. (2015) reports an increase in waste and contaminated land, a decrease in air quality, an increased number of heart attacks, and decreased crime rates. Tertiary education enrollment rates suffered a reduction of 28% in 2011 compared to 2010. Table 2 compares the Maule and Canterbury earthquakes in terms of direct damage.

**Table 2: A comparison of each earthquake's direct damage**

|   | **Variable** | **Maule 2010** | **Canterbury 2011** |
|---|---|---|---|
| 1 | Magnitude (Richter) | 8.8 | 6.1 |
| 2 | Total deaths | 562 | 181 |
| 3 | Total affected* | 2,671,556 | 301,500 |
| 4 | Total damage (billion US$) | 30 | 15 |
| 5 | Losses as a % of national GDP | 18% | 10% |
| 6 | Reconstruction government spending (billion US$) | 8.41 | 10.14 |

Source: Rows 1-4: EM-DAT (2018). Row 5: Estimated by dividing total damage by GDP from the World Bank (2018). Row 6: Chile - Government reconstruction plan (2010). New Zealand -Budget policy statement (2013). * Total affected refers to the number of people requiring immediate assistance and may include displaced or evacuated people.

Similarly, the impact on the housing stock of Canterbury was huge, with the earthquake affecting almost three-quarters of the total. 150,000 homes were damaged, and the cost of residential property damage was estimated at NZ$ 13 billion (Parker and

Steenkamp, 2012). Insured losses of over NZ$ 30 billion and over 750 thousand insurance exposure claims were submitted to New Zealand's Earthquake Commission (EQC) for damage to buildings, land, and contents. Businesses were severely affected, with 64% closing with a median closure time of 16 days. 11% of companies closed permanently (Potter et al., 2015). Besides the direct shock to the capital stock, businesses were also disrupted by the damage to infrastructure such as roads and utilities. Demand for some sectors was also affected, such as construction, which increased, and tourism, which decreased.

4. **Method and data employed**

The SCM is an empirical method for comparing outcomes in case studies. Since the articles by Abadie and Gardeazabal (2003) and Abadie et al. (2010), the SCM has been widely used for studying several types of policy changes in different fields, such as economics, development, political science, and medicine. The SCM is a methodological approach suitable for causal inference in case studies with one treated unit and limited macroeconomic data (Abadie and Gardeazabal, 2003; Abadie, 2021; Athey and Imbens, 2017). SCM combines aspects of the matching and difference-in-difference techniques to facilitate counterfactual comparisons. Recently, SCM has been employed in a variety of fields, such as political science (Abadie et al., 2015), political economy (Abadie and Gardeazabal, 2003; Absher et al., 2020; Grier and Maynard, 2016), and development (Billmeier and Nannicini, 2013; Spruk, 2019).

      The virtue of this methodology is that it strengthens the robustness of comparative case studies and its ability to provide quantitative inferences (Abadie et al., 2015; Athey and Imbens, 2017). Considering the limited access to granular macro data on economic growth and development across countries, the feasibility of constructing methodologically sound counterfactuals is extremely limited. Thus, the selection of this method becomes justified. Using pre-treatment data, SCM creates a synthetic counterfactual, a weighted average of control donors with similar conditions. The synthetic control design tracks the pre-treatment

outcomes and matches the treated unit values of several indicator variables to several donors (Gilchrist et al., 2023).[1]

A robust synthetic control makes it possible to study the effect of a random treatment, such as a natural disaster, longitudinally from the moment the disaster hits (Rayamajhee et al., 2024). Afterward, permutation and robustness tests must be performed to confirm validity as a counterfactual. Linham et al. (2017) assess the impact of a tsunami that struck Hawaii in 1960 in the long term, comparing their results to those of the affected island as far as 15 years after the disaster. For our case studies, we are interested in constructing a synthetic control *at the regional level* for the regions affected by the earthquake in Chile in 2010 (Maule) and New Zealand in 2011 (Canterbury). The SCM allows to measure the effect of a localized treatment or intervention on a single unit, which could be any entity for which the researcher has disaggregated data (i.e., city, region, university, etc.), by synthesizing a control unit from a donor pool (i.e., a group of untreated entities). Because the synthetic unit is constructed from a group of controls, it serves as a counterfactual for the treated unit, which, in our case, is the region most affected by an earthquake.

Following Abadie and Gardeazabal (2003), Absher et al. (2020), and most recently, Abadie (2021), we choose a pool of "donor" regions that share similar conditions to the treatment period for the independent variable. Such conditions include culture, history, geography, education, language, structural economic similarities, and institutional framework (Absher et al., 2020). Since our focus is regional, we follow the approach of Abadie and Gardeazabal (2003) and use other regions from within the same countries unaffected by the treatment (see Table 4). We use annual regional-level panel data from 1985 up to the interventions in 2010 and 2011 to build reasonable counterfactuals. After that, we evaluate the post-treatment effect up until 2015, which is in line with the methodological recommendations of Abadie (2021). The data and sources are summarized in Table 3.

Additionally, all our data sources are obtained from central banks and government agencies. In the case of Chile, our data comes from the Regional Observatory of Chile (2018) and the Central Bank (2019). We employ slightly different variables to construct each synthetic control due to the way each country builds its data sets. Still, the dependent variable, GDP per capita, remains the same across all cases. This allows us to compare the impact of earthquakes between different cases.

---

[1] For an in-depth review and analysis of the synthetic control methods, its mathematics, best practices, and applications consult: Abadie (2021), Abadie et al., (2015), Gilchrist et al. (2023), and Ben-Michael et al. (2021).

**Table 3: Fiscal and institutional comparison between Chile and New Zealand**

| Variable | Chile | New Zealand | Source |
|---|---|---|---|
| GDP per capita – 2009 (1990 US$ PPP) | $13,210 | $18,843 | Maddison |
| Change in Public Spending – year of earthquake | 10.44% | 6.24% | World Bank |
| Government Debt as a % of GDP (2011) | 11.10% | 62.60% | World Bank |
| Government Deficit as a % of GDP – year of earthquake | -3.45%* | -4.01% | OECD |
| Institutional Variables – 2010: | Value | Value | |
| Size of Government | 7.91 | 4.91 | Fraser Institute |
| Legal System & Property Rights | 6.80 | 8.69 | Fraser Institute |
| Sound Money | 8.94 | 9.65 | Fraser Institute |
| Freedom to Trade Internationally | 8.25 | 8.67 | Fraser Institute |
| Regulation | 7.55 | 8.92 | Fraser Institute |

Note: The change in public spending is shown in each country during the year of the earthquake, 2010 for Chile and 2011 for New Zealand. All variables except public spending are shown for 2009. *Deficit data for Chile obtained from Díaz et al. (2016).

In the case of Chile, to construct the synthetic control for the 2010 earthquake, we use the following variables: GDP per capita (averaged between 2005-2009), the share of population with tertiary education (2008-2009), and all available sectoral shares of GDP: agriculture, fishing, mining, manufacturing, energy, construction, retail, transport, financial, housing, personal, and public sector. We take the average values of all sectoral shares from 2005 to 2009. Similarly, for the NZ case, we take all economic data from Stats NZ (2018), the public service department responsible for collecting economic data. In addition to GDP per capita, the following sectoral GDP shares were used: agriculture, administration, construction, education, financial, food, health, information, manufacturing, occupation, professional, public services, rental, retail, transport, wholesale, and others. We average all economic activity data between 2006 and 2010—the five years before the earthquake. As a measure of human capital, the percentage of the population with a tertiary educational level is taken from Education Counts NZ and averaged between 2007 and 2010, since 2007 is the year for which the data starts.

The literature suggests that it is expected that a country with more effective regulation and better institutions will spend its resources more efficiently, especially in times of natural disasters (Acemoglu et al., 2014; Toya and Skidmore, 2007; Felbermayr and Gröschl, 2014). Also, it is more likely that a government-financed reconstruction program will impact GDP growth if it scores higher in those metrics (Barone and Mocetti, 2014; Díaz and Larroulet, 2021). The above suggests that a budget balance during the earthquake will also be relevant since the higher the government debt, the more difficult it will be to finance reconstruction efforts. High levels of government debt are costly since they make credit ratings fall and could decrease long-term growth in GDP per capita (Checherita-Westphal and Rother, 2012; Salmon, 2021). These variables are, therefore, incorporated in our synthetic experiments.

Finally, concerning the pool of 'donors' for each country, In the case of Chile, the donor pool will consist of Chilean regions that were not affected or were relatively unaffected by the earthquake. This allows for 11 out of the 12 regions available besides the Maule region since one of the other regions (the Bio-Bio region) was also significantly affected by the disaster and is therefore taken out of the donor pool. The regions considered come from the regional denomination from 1981 to 2007 since data were available based on that denomination. Available data include Tarapaca, Antofagasta, Atacama, Coquimbo, Valparaiso, Libertador Bernardo O'Higgins, Maule, Bio-Bio, Araucania, Los Lagos, Aysén del General Carlos Ibáñez del Campo, Magallanes, the Chilean Antarctic, and the Metropolitan region of Santiago.

In the case of New Zealand (NZ), the earthquake was of much lower magnitude, which is why only the Canterbury region was severely affected. Therefore, the other 14 regions of NZ are part of the donor pool. The regions are the ones reported by Stats NZ. They include Northland, Auckland, Waikato, Bay of Plenty, Gisborne, Hawke's Bay, Taranaki, Manawatu-Wanganui, Wellington, Tasmannelson, Marlborough, Westcoast, Canterbury, Otago, and Southland. The results of the weights of the different 'region donors' in the synthetic models can be seen in Table 4.

**Table 4: Synthetic weights and mean squared prediction error (RMSPE)**

|  | **Maule 2010** | **Canterbury 2011** |
|---|---|---|
| MSPE | 3.61E-01 | 8.50E-04 |
| MSPE ratio | 7.55 | 186.60 |
| Country Weights | Valparaíso<br>**(0.022)** | Auckland<br>**(0.3)** |
|  | Libertador Bernardo O'Higgins<br>**(0.063)** | Hawke's Bay<br>**(0.206)** |
|  | Araucanía<br>**(0.25)** | Manawatu-Wanganui<br>**(0.133)** |
|  | De Los Lagos<br>**(0.66)** | Tasman-Nelson<br>**(0.053)** |
|  | - | Marlborough<br>**(0.09)** |
|  | - | West Coast<br>**(0.218)** |

Note: RMSPE stands for root mean squared prediction error, and RMSPE ratio is the ratio of the RMSPE after and before the treatment. The table only shows the regions that reported positive weights for each region's different synthetic control models.

## 5. Synthetic control results: Chile (2010) and New Zealand (2011)

The SCM was implemented in R using the "synth package" developed by Abadie et al. (2011). The method is implemented at the regional level, taking the most affected region in each country as the treated unit. Now, we are ready to perform our synthetic control experiment. Figure 1 plots the real GDP per capita of the affected regions (solid lines) vis-à-vis their counterfactuals (dotted lines). Figure 1 displays actual and synthetic Chile's and New Zealand's real GDP per capita trajectory from 1985-2015.

**Figure 1. Synthetic controls and placebo tests for Maule and Canterbury**

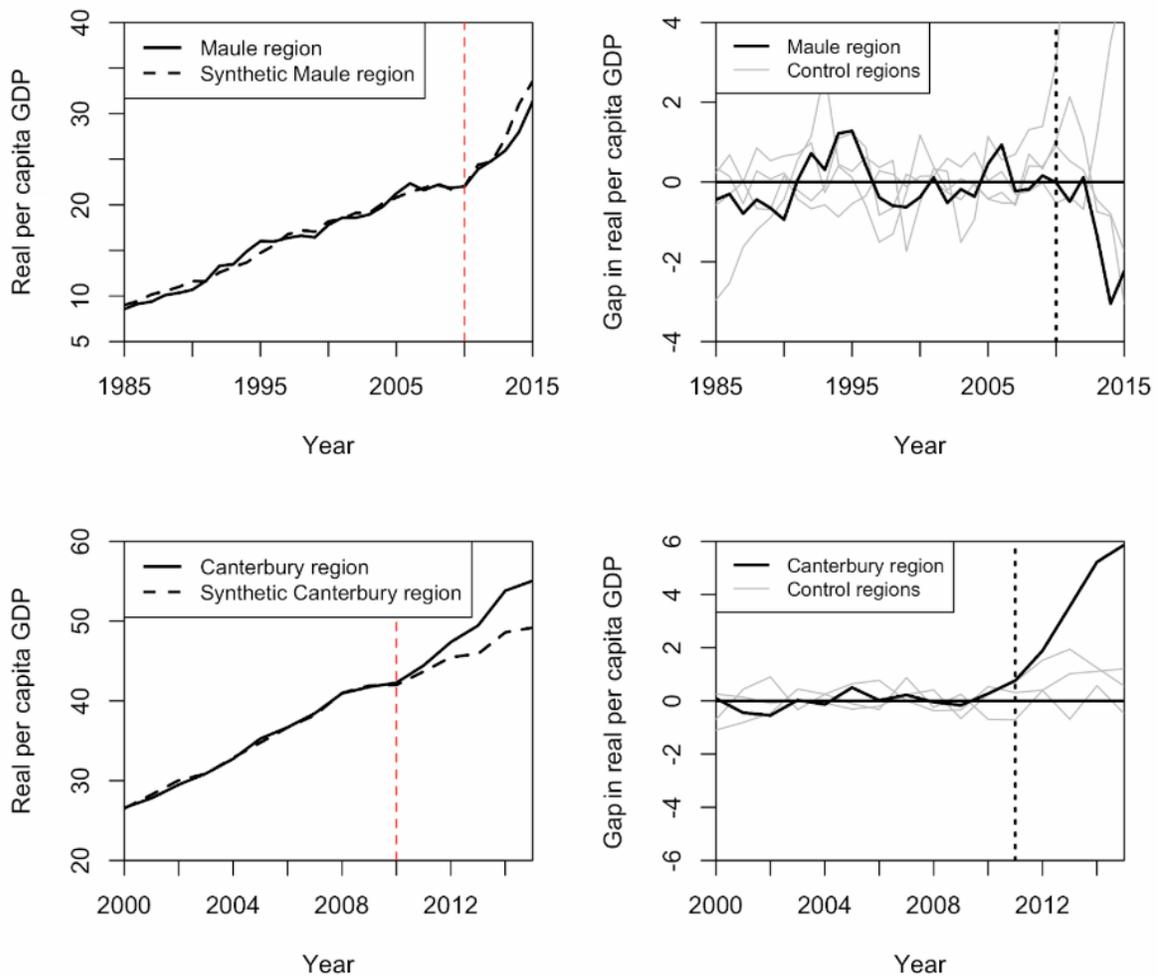

Note: The GDP per capita units for the Canterbury figure are thousands of current New Zealand dollars. The GDP per capita units for the Maule figure are $10^5$ Chilean pesos of 2003.

As depicted in Figure 1, synthetic Chile and NZ provide excellent pre-treatment tracking of actual Chile and NZ during the 25-year and 24-year pre-treatment periods, respectively, especially in the years closer to the intervention. For the synthetic control to be a representative counterfactual of the region of interest after the intervention, robustness tests must be performed to ensure that the treated unit experiences a *unique* treatment. This is accomplished with placebo tests, which consist of applying the control method to every other unit in the sample to check if others produce a gap like, or more significant than, the treated unit after the treatment starts. Placebo tests (on the right side of the figure) are thus provided for robustness.

For Canterbury, surprisingly, we find positive effects on GDP per capita in the year of the earthquake and every year after it for four consecutive years. In 2014, three years after the event, the synthetic counterfactual of Canterbury was 9.81% *lower* than the actual value. The case of NZ then is interesting and counterintuitive: unlike what happened in Kobe, Japan, in 1995 that experienced a persistent 12% decrease in GDP per capita after the earthquake (duPont IV and Noy, 2015), Canterbury instead experienced a persistent 9.81% *increase* in GDP per capita. This result is akin to what BM found for the Friuli region in Italy. In the case of Chile, in contrast, the effect is the opposite and closer to the general one predicted by the literature, and the synthetic counterfactual has a real GDP per capita *higher* than the actual GDP of the region. In 2013, three years after the earthquake, the synthetic Maule region had a GDP per capita that was 5.15% higher than without the quake compared to what happened under the presence of the intervention. For Chile, we identify a decline in per capita GDP attributable to the earthquake, which is persistent, long-term, and observable even five years after the quake. GDP per capita for 2015 was lower (% decrease) than it would have been had the earthquake not occurred.

According to our placebos results (see Figure 1), both earthquakes act as a treatment that affects future GDP per capita since only a single other region shows a larger gap after the earthquake in the case of Chile, and no region shows a larger gap in the case of New Zealand. The ratio of control units showing a larger gap than the treated units over the total number of units is considered a symbolic p-value, representing the probability that a non-treated unit will show an effect as significant as the treated unit. In our case, this p-value would be 0 for the analysis of the Canterbury region in New Zealand and close to 7% for the Maule region in Chile. This suggests that our results are robust and significant.

Notably, according to our empirical analysis, synthetic Canterbury has a higher GDP per capita than its real counterpart with the presence of the earthquake (the real Canterbury). However, this result is an indirect effect of the disaster and says nothing about the aggregate impact of the event on the national economy. Considering the government resources spent, the net effect might still be negative, at least in the long term. With the synthetic control, we can examine the recovery process from both earthquakes in the short and the long term, and the effect is persistent in time in the period considered for both countries after the intervention.

Given the large amount of government spending for reconstruction purposes in Canterbury vis-à-vis Maule (keeping also in mind the differences in magnitude and damage between the two events), this is likely an important factor that explains the differences in *ex-post-treatment* growth between these regions. The total amount pledged by New Zealand for reconstruction was $10.14 billion US, while the amount pledged for Chile was $8.41 billion US, according to each government reconstruction plan; these relief packages *correspond to 67.6% and 28.0% of each earthquake reported damage*, respectively. In other words, NZ pledged more than double the amount of government relief and reconstruction efforts compared to Chile if we consider the difference in reported damage. The literature on natural disasters suggests that large amounts of government reconstruction aid could be relevant in mitigating the adverse economic effects of earthquakes (Barone and Mocetti, 2014; Lian et al., 2022). A related possibility is that Canterbury might have experienced also the 'productivity effect', which is the possibility that "disasters might have positive economic consequences, through the accelerated replacement of capital" (Hallegate and Dumas, 2009, 777). Thus, the productivity effect could have had "a significant impact on the production level since, when effective, it can cancel the long-run losses due to disasters" (ibid.).

Finally, institutional quality and the higher regard for property rights in New Zealand could explain some of the differences regarding the impact of the earthquake in comparison to Chile, as there were several cases of looting and disrespect for the rule of law after the disaster in the Maule region (Dussaillant and Guzman, 2014), while nothing to the same degree was observed in New Zealand. The literature also suggests that institutional quality, trade openness, higher income, and higher educational attainment are crucial to mitigate the adverse effects of natural disasters (Toya and Skidmore, 2007; Felbermayr and Gröschl, 2014). The case of New Zealand, when compared to Chile, seems to confirm this.

### *5.1. Impact on different economic sectors*

In this subsection, we examine what happened with multiple economic sectors in Chile and New Zealand by disaggregating those economies and focusing on sector and industrial data. Considering that the GDP per capita of the Canterbury region increased because of the earthquake, according to our synthetic control analysis, it should be expected that at least one sector of the NZ economy experienced an important increase after 2010. We will test this hypothesis using the SCM for different economic sectors.

#### *5.1.1 Economic sectors' analysis for Canterbury*

It is expected that considering the toll that the earthquake had on residential property in Canterbury (Miller, 2014), the construction sector would have experienced a significant increase in demand, which would, in turn, have increased the sector's GDP. As shown in Figure 2, GDP in the construction sector increased dramatically after the earthquake, a result that is robust according to our placebos. According to the synthetic control, the GDP of the actual economic sector (solid line) increased close to 1000 $NZ two years after the disaster compared to the counterfactual one without the intervention (dotted line), which is roughly 40% of the region's entire construction sector GDP. The fit of the synthetic control is high, with a root mean squared prediction error (RMSPE) of 48.9 $NZ. The destruction included roads and other infrastructure not limited to residential properties. About 1.600 buildings had to be wholly or partly demolished (Brownlee, 2012).

On the other hand, Canterbury's agricultural sector was deemed largely unaffected, according to Parker and Steenkamp (2012). Our results contradict that assessment: We perform the SCM for that sector and show the results in Figure 3, which suggests that the real agriculture sector *increased its production* after the earthquake in comparison to the synthetic control, achieving a level 25% higher than the synthetic control, one year after the earthquake. One of the biggest challenges for the agriculture sector was the disruption in essential services and roads. Some dairy farms reported problems with electricity disruption and structural damage, while arable farms had more trouble with the interruption of water services. Fortunately, the timing of the earthquake was so that irrigation was unnecessary (Whitman et al., 2012).

**Figure 2. Synthetic control and placebo for the construction sector (Canterbury)**

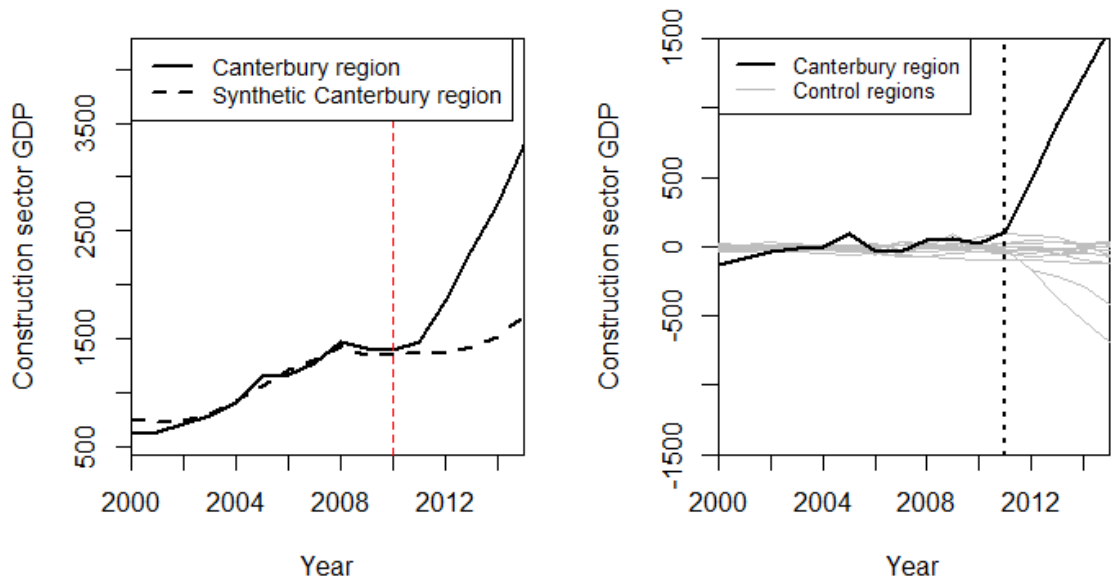

Note: units are thousands of NZ Dollars.

We also study the impact on Canterbury's financial sector GDP, finding that the earthquake had negative and persistent effects as commonly expected. In 2012, the GDP of the sector declined close to 20% compared to 2011, and the synthetic control shows that it would have stayed around 2011 levels had the earthquake *not occurred*. The trajectory of the synthetic control of the financial sector's GDP closely follows the actual one, and the result is robust according to the placebo test since no other region shows a gap close to the one of Canterbury compared to their synthetic controls. Unlike the positive results on economic activity obtained in both the actual construction and agricultural sectors with the intervention, our analysis shows that the financial sector experienced a severe decline compared to counterfactuals.

**Figure 3. Synthetic control and placebo test for the agriculture sector (Canterbury)**

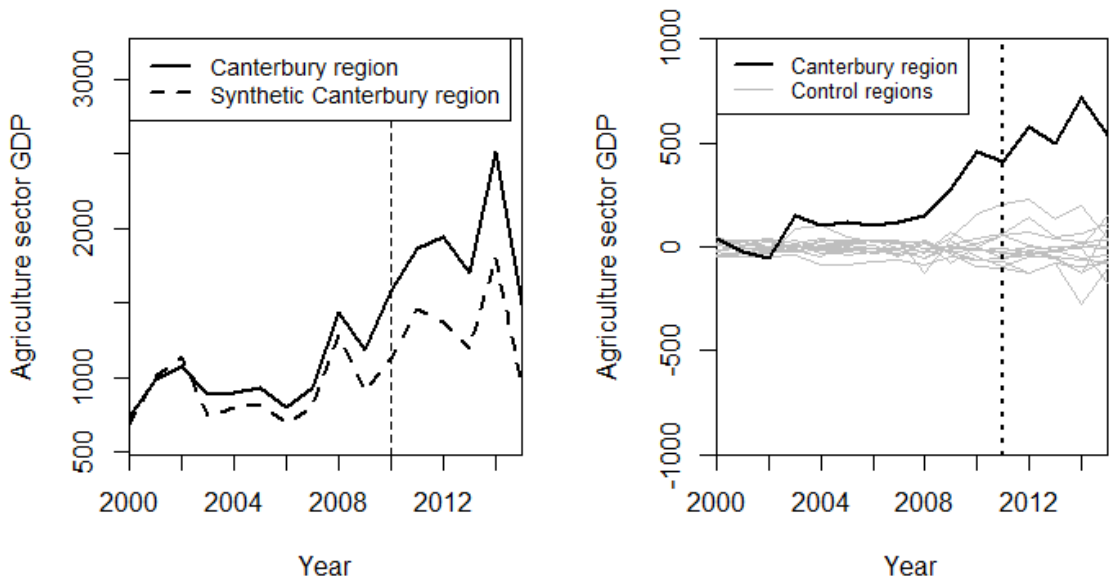

Note: units are in NZ Dollars.

**Figure 4. Synthetic control and placebo test for the financial sector (Canterbury)**

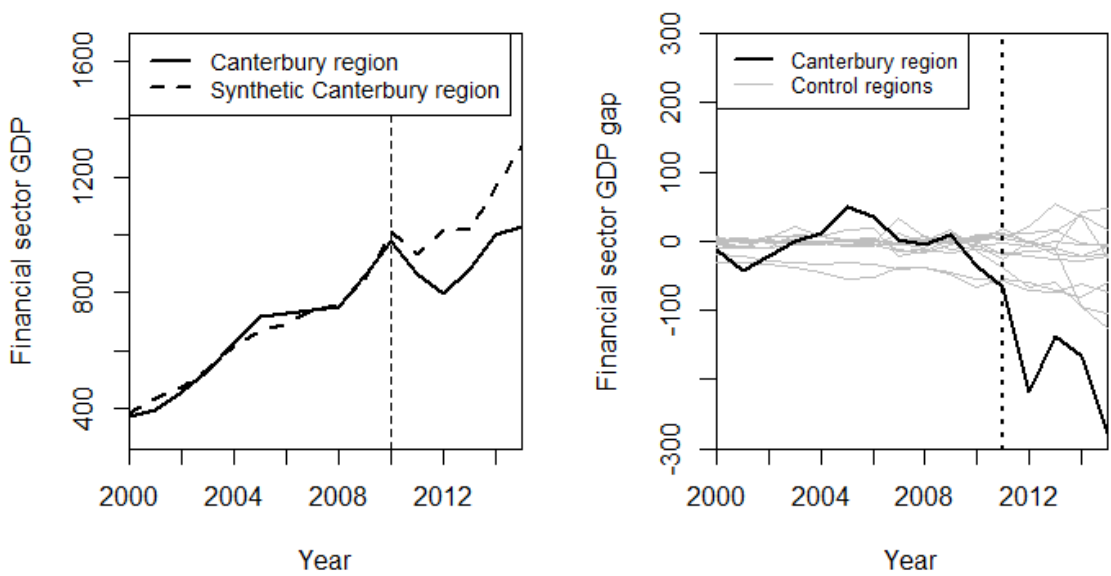

Note: units are in NZ Dollars.

After the earthquake, extensive insurance helped limit losses in the banking sector. Because of the high amount of insurance claims, the government provided a financial support package to an insurer of residential property, while some insurers required capital injections from their parent companies. Extensive reinsurance was also abroad, which helped reduce the impact on New Zealand's insurance companies (Parker and Steenkamp, 2012).

The manufacturing sector, in contrast, was not heavily affected by the earthquake since most of its activity was located outside the areas that sustained most of the damage (Parker and Steenkamp, 2012). According to the synthetic control analysis (see Figure 5), the earthquake's impact was mildly positive and relatively low, less than 200 $NZ higher than the synthetic counterfactual one year after the quake and slightly decreasing in 2015, suggesting no substantial long-term effects for this sector.

**Figure 5. Synthetic control and placebo test for the manufacturing sector (Canterbury)**

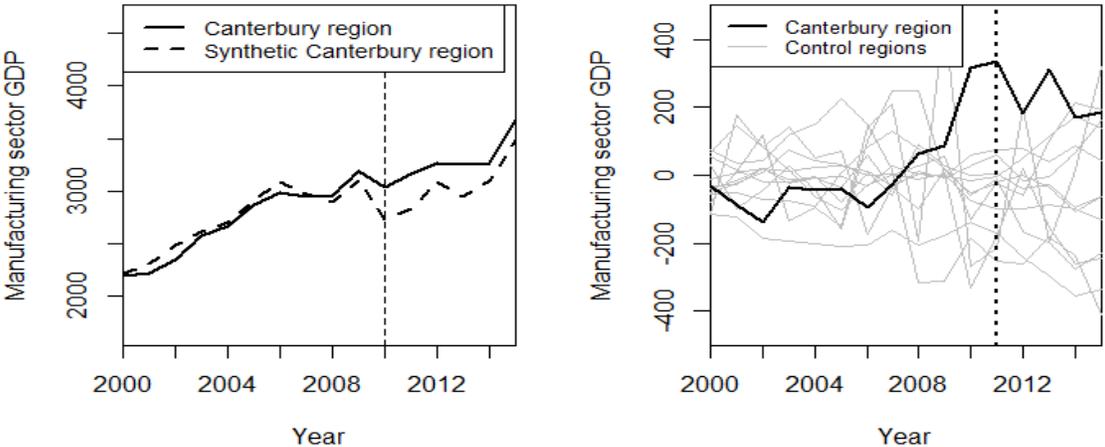

Note: units are in NZ Dollars.

Finally, the accommodation sector is expected to have declined after the earthquake similar to the financial sector. Usually, after natural disasters, tourism declines temporarily, and some hotels might be even forced to close. Canterbury was no exception, and even though the sector did not experience a severe decline, it did stop growing, as Figure 6 attests. According to the synthetic control (see Figure 6), the trajectory of the accommodation and food sector of the counterfactual suggests that the industry would have continued to grow if the earthquake *had not* occurred, reaching a level 14.2% higher three years after the disaster.

**Figure 6. Synthetic control and placebo test for the accommodation sector (Canterbury)**

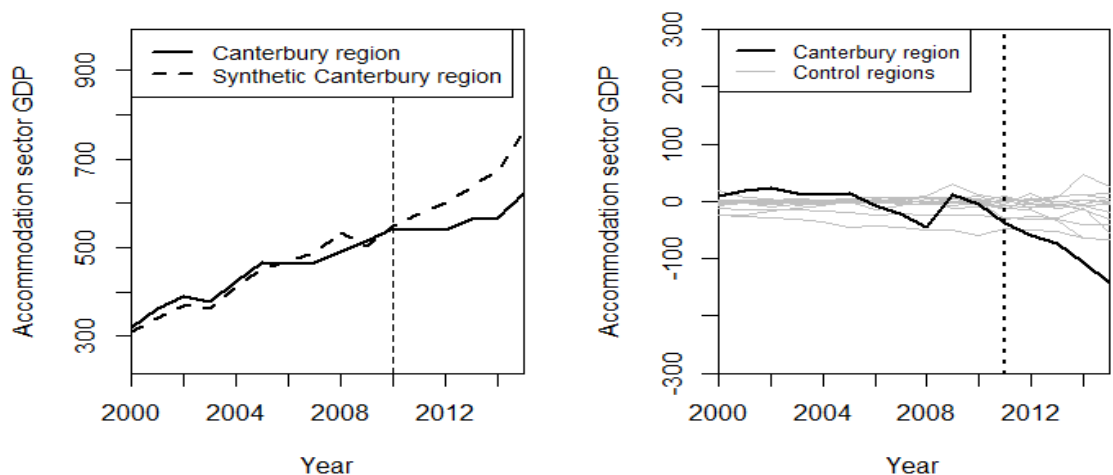

Note: units are in NZ Dollars.

*5.1.2 Economic sectors' analysis for Maule*

We also study what would have happened in Chile at the sectoral level using the SCM. Starting with the construction sector (See Figure 7), our experiments show that it experienced a sharp increase immediately following the year of the earthquake, reaching a value of 28.6% higher in 2012 (solid line) than the synthetic counterfactual without the earthquake (dotted line). Still, this effect is not permanent and withers away in the long run.

Additionally, the agricultural (in combination with the silvicultural) sector also experienced an increase, or more precisely, *declines less* than it would have without the disaster, since the synthetic control is persistently lower than the actual sectoral GDP trajectory (see Figure 8). According to both placebo tests, the results are robust, with no other units experiencing a more significant gap in the construction sector and just one other experiencing a more significant gap in the agricultural-silvicultural industry by the last two periods, representing a p-value of 7%.

**Figure 7. Synthetic control and placebo test for the construction sector (Maule)**

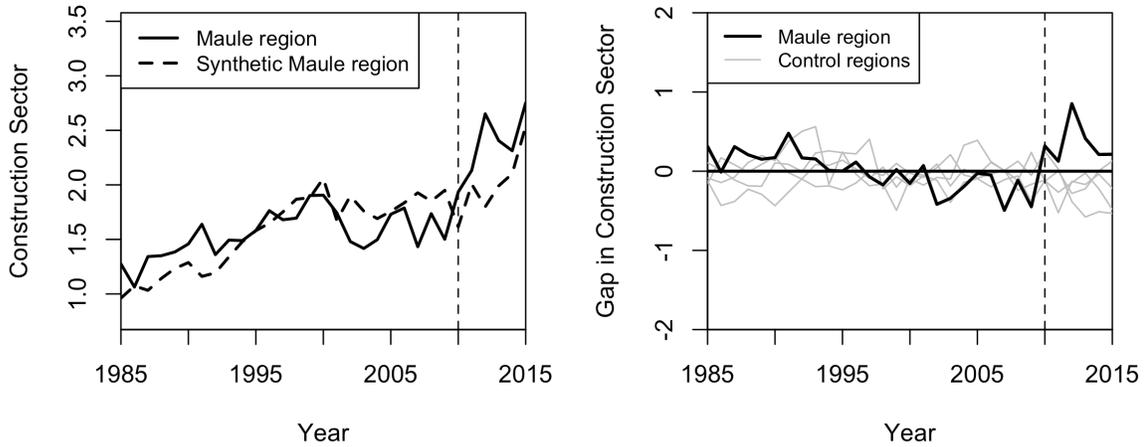

Note: Units are 105 Chilean pesos of 2003.

**Figure 8. Synthetic control and placebo test for the agricultural sector (Maule)**

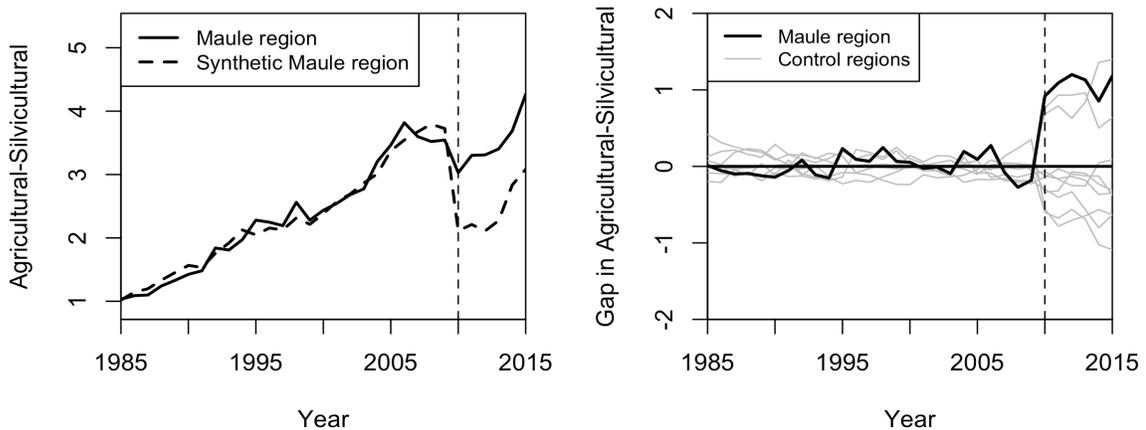

Note: Units are 105 Chilean pesos of 2003.

The financial sector experienced minimum changes after the earthquake in the Maule region (see Figure 9), with the synthetic counterfactual closely following the real sector's GDP during the entire period. Although the total losses from the earthquake were estimated at 30 billion $US, the amount insured covered about 8 billion $US (Saragoni, 2011), or 26.7% of the estimated losses.

**Figure 9. Synthetic control and placebo test the financial sector (Maule)**

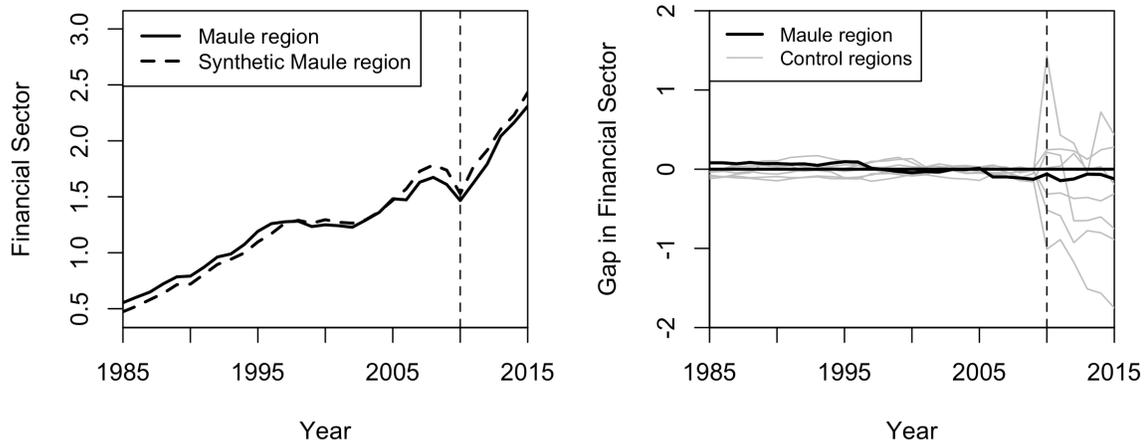

Note: Units are 105 Chilean pesos of 2003.

**Figure 10. Synthetic control and placebo test for the manufacturing sector (Maule)**

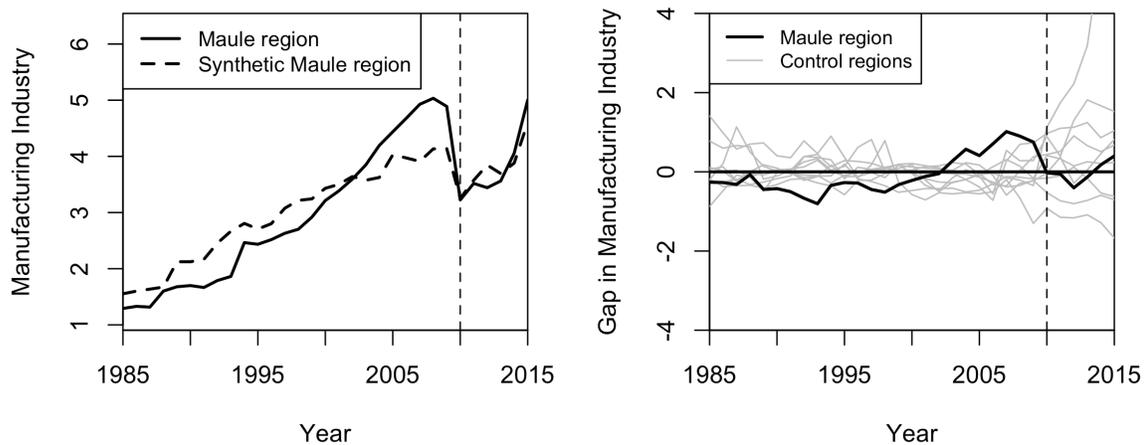

Note: Units are 105 Chilean pesos of 2003.

Additionally, the GDP of the manufacturing sector does not appear to be affected at all in our analysis. However, the synthetic counterfactual (see Figure 10) does not closely follow the GDP of the Maule region, providing a defective fit. It is also not possible to conclude that our results are robust for this sector since multiple regions in the robustness test experience more significant gaps after the event as seen from the placebo tests.

## 6. Conclusions

Following the incipient literature applying synthetic controls on natural disasters, we performed case studies that determined the impact of the recent earthquakes in Chile and New Zealand at the regional level. From these two largely neglected experiences closely related to each other in terms of distance and magnitude, we have shed light on the plausible economic consequences of natural disasters such as earthquakes on the wealth and development of regions and how their effects could be mitigated through government policy responses. The economic literature on natural disasters has largely neglected these two cases despite being relevant earthquakes that have occurred during the past decades. Illuminating these cases and assessing their dissimilar impact through the SCM has been the core contribution of this work, closing a gap in the literature.

Additionally, by focusing on a regional level, we have been able to study these two cases on a more granular level, obtaining results about how these disasters have affected both places differently in relation to their policy responses and institutions. Using the earthquakes as treatments when applying synthetic controls, and by using reasonable counterfactuals based on regions in the same countries that were not affected (or were slightly affected) by the earthquakes, we found: First, that the actual Canterbury region achieved *higher levels* of GDP per capita in the years after the disaster than it would have achieved in the absence of the disaster (the counterfactual), and this effect seems to be persistent. The positive role of the earthquake in Canterbury is consistent with the literature that suggests that both disaster relief and financial aid for the reconstruction could play a positive role in capital accumulation and boosting economic activity. Additionally, the high-quality (inclusive) institutions of New Zealand could also help explain the results. Second, in the case of Chile in 2010, the Maule region experienced a *decline* in GDP per capita due to the earthquake when compared to its counterfactual. The negative role of the earthquake in Maule could be explained by both the meager disaster relief and financial aid programs for the reconstruction pledged by Chile (compared to NZ) and the quality of its institutions. The institutional quality and the higher regard for property rights in New Zealand could explain some of the differences regarding the impact of the earthquake compared to Chile.

Our findings are consistent with Barone and Mocetti (2014), making a contribution to the incipient literature on the economic consequences of disasters using the SCM. Our results provide additional support for Noy's (2009) findings, suggesting that the severity of the macroeconomic output damage depends largely on the level of development: "developing countries … face much larger output declines following a disaster of similar relative magnitude than do developed countries or bigger economies". Countries with better institutions, higher per capita income, higher degree of openness to trade, and higher levels of government spending withstand disasters much better, preventing spillovers into the macro-economy as the cases studied highlight.

In terms of magnitudes, the GDP per capita of the synthetic control of the Canterbury region was 9.81% lower than the region's actual value three years after the earthquake of 2011, while the actual Maule region saw a decline of 5.15% after the disaster against its counterfactual, also three years after. All our results are robust according to our placebo tests. Considering the significant government spending regarding reconstruction, US$ 8.4 billion for Chile after the 2010 earthquake and US$ 10.1 billion for New Zealand (accounting for a percentage of earthquake damage of 28% and 66%, respectively), it is likely that the positive impact on regional growth is mainly accounted for by the reconstruction-led boom after each earthquake and that this effect might be compounded throughout the role that better institutions play in mitigating natural disasters.